\documentclass{article}
\usepackage[T1]{fontenc} 
\usepackage[utf8]{inputenc} 
\usepackage{ismir,amsmath,cite,url}
\usepackage{graphicx}
\usepackage{color}
\usepackage{comment}
\usepackage{commath}
\usepackage{booktabs}

\usepackage{float}
\usepackage{siunitx}
\usepackage{lineno}

\title{A Deep Learning Based Analysis-Synthesis Framework For Unison Singing}

\multauthor
{Pritish Chandna$^1$ \hspace{1cm} Helena Cuesta$^1$ \hspace{1cm} Emilia G\'omez$^{2,1}$} { \\
  $^1$ Music Technology Group, Universitat Pompeu Fabra, Barcelona\\
$^2$ European Commission, Joint Research Centre, Seville\\
{\tt\small \{pritish.chandna, helena.cuesta, emilia.gomez\}@upf.edu}
}
\def\authorname{P. Chandna, H. Cuesta and E. G\'omez}

\begin{document}

\maketitle
\begin{abstract}
Unison singing is the name given to an ensemble of singers simultaneously singing the same melody and lyrics. While each individual singer in a unison sings the same principle melody, there are slight timing and pitch deviations between the singers, which, along with the ensemble of timbres, give the listener a perceived sense of "unison". In this paper, we present a study of unison singing in the context of choirs; utilising some recently proposed deep-learning based methodologies, we analyse the fundamental frequency (F$0$) distribution of the individual singers in recordings of unison mixtures. Based on the analysis, we propose a system for synthesising a unison signal from an \emph{a cappella} input and a single voice prototype representative of a unison mixture. We use subjective listening tests to evaluate perceptual factors of our proposed system for synthesis, including quality, adherence to the melody as well the degree of perceived unison.
\end{abstract}
\section{Introduction}\label{sec:introduction}
Throughout history, singing has been an important cultural activity for humans, serving for propagation of beliefs and ideas amongst the masses as well as for social entertainment. The social aspect led to gatherings of people singing in a group, which evolved into polyphonic ensemble singing with multiple voices singing counterpoint melodies in complex harmonies. A group of people singing in such an ensemble is commonly termed as a choir and the focus of our study is on one setting of such choirs consisting of four voices known as Soprano, Alto, Tenor and Bass (SATB). Each voice within an SATB ensemble has its own function and melodic range in the whole. SATB is one of the most widely studied, documented, and practiced forms of choirs with numerous dedicated conservatories across Europe, highlighting the cultural importance of the art form. Within each of the SATB voices, it is common to have multiple singers of similar vocal range singing the same melody simultaneously, in a form known as unison singing. While all the singers in a unison sing the same melody, it is impossible for a group of two or more people to perfectly synchronize and sing the exact same pitch line. Each singer has their own natural micro-deviations, both in terms of timing and pitch, from the prescribed score and their own distinct timbre. The combination of micro-deviations and the ensemble of timbres leads to the perception of unison, wherein several singers are perceived to be singing a single pitch contour~\cite{Ternstrom91_PerceptualEvalVoice}, and is the main focus of our study. 

Pitch and fundamental frequency (F$0$) are related but not equivalent terms. While the F$0$ generally refers to the physical frequency of vibrations of the vocal folds for a singing voice signal, pitch refers to an abstract perceptual concept which has been found to be closely correlated to the F$0$. Frequency is usually measured in Hertz, representing the number of cycles of a periodic signal per second, whereas pitch is described in terms of perceptual units like \textit{cents}. The cent is a unit defined on a logarithmic scale, as a measure of the ratio between the frequency in Hertz and a base frequency, commonly chosen to be \SI{440}{\hertz}, as shown in Equation~\ref{eq:posen2}.
\begin{equation}
f_{cents} = 1200\cdot log_{2}\frac{f_{hertz}}{440}
\label{eq:posen2}
\end{equation}
Thus defined, the cent is correlated to the perceptually relevant musical unit of an equally tempered semitone. Specifically, one semitone spans \SI{100}{cents}. Examined individually, the pitch of the singers in a unison can be represented by the F$0$ of each individual singer's vocal signal, this can be tested by synthesising a time-varying sinusoid with the frequency of the signal. However, when the individual signals are added the resultant pitch is not merely the sum of F$0$ value, and the methodology of synthesising a sum of sinusoids as with single singers fails to produce the same perceptual result due to physical phenomena such as beating, among others. Past studies have utilized artificial unison mixes created by the use of a vowel only singing voice synthesizer to study the perception of a single pitch contour in a unison~\cite{Ternstrom91_PerceptualEvalVoice}. Other areas of past research related to unison singing include single voice to unison synthesis models, based on creating voice clones with variations in the input~\cite{schnell2000synthesizing, Bonada2005_SoloToUnison_AES}. 
 Fuelled by the deep learning revolution, singing voice synthesizers have evolved over the last few years, allowing us to take a step further both into exploring the perception of unison and into effective solo voice to unison synthesis. We build on the work done by Ternstr\"om~\cite{Ternstrom91_PerceptualEvalVoice} by leveraging recently proposed synthesis methodologies to synthesize a single voice prototype representing the melodic and linguistic content of a unison mixture. This allows use to further test the hypothesis of a single F$0$ contour representative of the perceived pitch of a unison via subjective listening tests. We also verify the author's findings by analysing a set of real recordings of unison singing. 
 In addition, we propose a methodology combining previous research and recently developed techniques to synthesize a unison mixture from a single voice input. We follow the basic methodology used by Schnell et al.~\cite{schnell2000synthesizing} to create voice clones with variations in three aspects; pitch, timing and timbre, and use perceptual evaluation tests to evaluate the effect of each of these parameters on the perception of the sense of a unison.
 
The rest of the paper is structured as follows. Section~\ref{sec:sota} discusses previous works pertinent to our study. We then present the analysis of the choir recordings in Section~\ref{sec:analysis}, including a description of the dataset of choir recordings, the methodology used for the analysis and the results of the analysis. The synthesis methodology we use for synthesizing voice clones and the single voice prototype of the unison mixture is described in Section~\ref{sec:method}. Section~\ref{sec:evaluation} presents the perceptual evaluation methodology used and the results of the perceptual tests. Finally, we present a discussion on our findings in the analysis of the choir and the perceptual evaluation of the synthesis in Section \ref{sec:conclusions}. 

\section{Related work}\label{sec:sota}
We divide the description of related works into three sections: past studies into the analysis of the perception of unison, previous works on synthesising unison mixtures from choirs and recently proposed deep-learning based methodologies what we will use for analysis and synthesis.
\subsection{Analysis Of Unison Perception}
The perception of pitch dispersion has previously been studied in~\cite{Ternstrom91_PerceptualEvalVoice}, wherein the author used synthesized singing voice stimuli to investigate the preferences of expert listeners in unisons. In the study, pitch dispersion is defined as the bandwidth of the fundamental frequency and its harmonic partials across individual singers in a unison. It is suggested that this dispersion is related to the \emph{flutter}---small variations in F$0$ that are too fast to be perceived as pitch variations. The concept of pitch \emph{scatter} is presented in the study as the standard deviation over voices in the mean F$0$: the average F$0$ computed over the duration of each tone of a song. The study concludes that a \emph{scatter} of \SIrange{0}{5}{cents} was preferred by the participants while a \emph{scatter} of \SIrange{5}{14}{cents} was seen as the limit of tolerance before dissonance was reported. In addition, the author also highlights several differences between solo and ensemble singing. For instance, a single performer produces tones with well-defined properties: pitch, timing, loudness, timbre, while an ensemble of performers produces sounds with statistical distributions of each of these properties. 

A similar method for modelling \emph{scatter} in choir sections was presented by Cuesta et al.~\cite{CuestaGML18_ChoirIntonation_ICMPC} using small windows to compute the standard deviation between individual F$0$s in the unison. This study used real recordings of choral singing instead of synthesised stimuli, presenting a mathematical model for dispersion rather than a perceptual evaluation. For the dataset used in the research, F$0$ (or pitch) dispersion was found to be in the range of \SIrange{20}{30}{cents} for all SATB voice sections, being slightly larger in the Bass. 

Another recent study focused on the analysis of F$0$ in vocal music is work by Weiss et al.~\cite{WeissSRM19_IntonationLocusIste}, where the authors proposed an approach to measure intonation quality of choir recordings. They create an ideal \num{12}-tone equal temperament grid, and then calculate the deviation of each F$0$ and its partials to their theoretical position in the grid. The overall deviation is computed as a weighted sum of each partial's deviation. This method enables the analysis of the overall intonation of a full choir recording, but does not account for the deviations within sections of the choir. 

\subsection{Unison Synthesis}
Signal processing techniques have previously been utilised to synthesize choir unison by adding several clones of a monophonic \emph{a cappella} signal with uncorrelated pitch, timing, and timbre deviations. Most particularly, Pitch Synchronous Overlap Add (PSOLA) methods~\cite{schnell2000synthesizing} have been exploited as an analysis-synthesis framework to decompose the vocal signal into a set of constituent waveforms representing successive pitch periods of the signal. Pitch and timing deviations are added to the vocal signal using time stretching and pitch shifting techniques to create voice clones, which are combined to form the output unison signal. 

Other proposed methodologies for creating a unison output from an \emph{a cappella} signal include morphing the spectral and pitch components of the vowels of the input signal as in~\cite{Bonada2005_SoloToUnison_AES}. The methodology's effectiveness is constrained to low tempo inputs . Random modulation of  beating partials to create a choral effect~\cite{dolson1983tracking} has also been used.

\subsection{Deep Learning For Analysis and Synthesis}
For our work, we build on the work done in~\cite{Ternstrom91_PerceptualEvalVoice} and~\cite{CuestaGML18_ChoirIntonation_ICMPC}, modelling the perceptual pitch contour of a unison mixture as a single F$0$ contour. To this end, we use a recently proposed Convolutional Representation for Pitch Estimation (CREPE) methodology~\cite{KimSLB18_CREPE_ICASSP} for extracting F$0$ contours from real world recordings of individual singers in a choir setting as well the F$0$ contour of unison mixture created by combining the individual voices. This methodology uses a series of convolutional operations on the waveform of the input signal and outputs a probability distribution over a discrete representation of the underlying F$0$ contour of the signal across a series of time-frames.


To synthesize the single voice prototype representing a unison mixture output and the voice clones for creating a unison mixture from a single voice input, we adapt the methodology proposed by Chandna et al.~\cite{chandna2020content}, which allows for the re-synthesis of a solo single voice from a musical mixture input via the underlying linguistic features. This methodology builds on the idea of re-synthesizing a vocal signal from a musical mixture by estimating the parameters of a vocoder synthesizer \cite{ChandnaBBG2019_Vocoder} and uses an encoder built of a bank of bi-directional long short-term memory (LSTM) recurrent neural networks (RNNs) to estimate a continuous representation of the underlying linguistic features present in the input mixture signal. The continuous representation is singer-independent and language agnostic, and was initially proposed for zero-shot voice conversion via an autoencoder network~\cite{qian2019autovc}. The linguistic features can then be used to generate the spectral envelope of the vocal signal in the mixture, providing the singer identity. The authors of~\cite{chandna2020content} proposed two decoders for this process, a Singer Dependent Network (SDN) which takes the singer identity as a one-hot vector, and a Singer Independent Network (SIN) which intrinsically learns the singer identity from the given input. The spectral envelope is then combined with the F$0$, extracted via an external algorithm, to synthesize the vocal signal. While the original framework was proposed for extracting a singing voice from a pop/rock musical mixture, we adapt the SDN network for synthesising a unison mixture from an \emph{a cappella} input and the SIN network for synthesizing an \emph{a cappella} singing voice from a unison mixture. The adaptations we apply are described in Section \ref{sec:method}. The SIN and SDN models~\cite{chandna2020content} were trained on a proprietary dataset with \num{12} hours of data comprising \num{205} songs by \num{45} male and female singers, and we have obtained a copy of the trained model with permission from the relevant authorities for our study. 

\section{Analysis Of Choir Recordings}
\label{sec:analysis}
We analyse the  variations between  individual singers in a unison in terms of variance in pitch and timing. Below, we present the dataset that we use in our analysis, followed by the methodology used for analysis, and finally the results of our analysis.
\subsection{Datasets}
We analyse the Choral Singing Dataset (CSD)~\cite{CuestaGML18_ChoirIntonation_ICMPC}, which includes monophonic recordings of \num{3} choral pieces: \textit{Ni\~no Dios d’Amor Herido}, written by Francisco Guerrero, \textit{Locus Iste}, written by Anton Bruckner, and \textit{El Rossinyol}, a Catalan popular song. There are \num{16} different singers for each song with \num{4} singers for each of the four parts; Soprano, Alto, Tenor and Bass. The dataset also includes manually corrected F$0$ annotations for each track. 
\subsection{Analysis Methodology }

To analyse inter-singer variance in \textbf{pitch}, the first step is the extraction of an F$0$ contour from a unison mixture of singers. We aim to study the behavior of a monophonic F$0$ extractor in such cases, assuming that we have a sufficiently balanced unison performance, where the contribution of each singer is similar in terms of volume and energy. 
To this end, we use  CREPE~\cite{KimSLB18_CREPE_ICASSP} to extract the fundamental frequency of the unison mixture created by summing and normalizing all corresponding individual singers in each vocal part of the recordings. This is termed as $EstF0_U$.

We then measure the resemblance of the estimated $EstF0_U$ to each of the manually annotated F$0$ tracks and to the mean $F0_m$\footnote{Note that the average F$0$ value has to be adjusted for timing differences between the individual singers. To this end, we define the average to be zero (unvoiced frame) if and only if all individual values for that frame are zero. for all other cases, the average is calculated only accounting for the non-zero values.}.
We use standard evaluation metrics for melody extraction including \emph{Raw Pitch Accuracy (RPA), Overall Accuracy (OA), Voicing Recall (VR)} and \emph{Voicing False Alarm (VFA)} between the $EstF0_U$,  the average the mean $F0_m$, and each individual singer curve, $GTF0_i$~\footnote{We use the \texttt{mir\_eval} library~\cite{RaffelMHSNLE14_mir_eval} for this evaluation, and we use a pitch tolerance of \SI{30}{cents}}

Once we have verified the accuracy of the extraction system, we build a statistical model for the individual contours in the unison, as suggested by~\cite{Ternstrom91_PerceptualEvalVoice}. In our model, the framewise F$0$ of an individual singer, $F0_i$, can be represented as a distribution of values around the mean $F0_m$ with a deviation of $F_{devi}$, as shown in Equation \ref{eq:model}

\begin{equation}\label{eq:model}
F0_i = F0_M + F_{devi}
\end{equation}

This equation also allows us to define the $F0_{i+1}$ of a singer in terms of the $F0_{i}$ of another singer in the unison as 

\begin{equation}\label{eq:model2}
\begin{split}
F0_{i+1} = F0_m + F_{devi+1}
\\F0_{i+1} - F0_{i} =  F_{devi+1} - F_{devi}
\\F0_{i+1} = F0_{i} + F_{devi+1} - F_{devi}
\\F0_{i+1} = F0_{i} + \Delta F0_s
\end{split}
\end{equation}
Where we define $\Delta F0_s$ as the inter-singer deviation, represented by Equation \ref{eq:deviation}. For each pair of singers in the unison, we compute the frame-wise difference between the corresponding F$0$ contours in cents. For this calculation, only frames with positive F$0$ values, also known as \textit{voiced} frames, were considered. We average these inter-singer deviations across time and songs, and obtain a single value for each group, i.e., SATB.

\begin{equation}\label{eq:deviation}
    \Delta F0_s = \frac{
    \displaystyle\sum_{i=1}^{n}\sum_{j=i+1}^{n}\abs{F0_i - F0_j}
    }{{n\choose 2}}
\end{equation}
where the sub-index $s$ indicates the choir section, $s\in [S, A, T, B]$, and $n$ is the number of singers. In our use case, $n=4$. 

To study \textbf{timing} deviations, we focus in the transitions from \textit{voiced} to \textit{unvoiced}, and vice-versa---where singers are not entirely in sync. We call these regions \emph{transition regions}, where some of the singers in the mixture are voiced and others are unvoiced, with a positive or zero F$0$. We measure the length of all the transition regions in every unison from the CSD, and then average across choir sections. 



%
\subsection{Analysis Results}
\label{sec:analysis:results}
\begin{figure}
 \centerline{
 \resizebox{0.95\columnwidth}{!}{
 \includegraphics[width=\columnwidth]{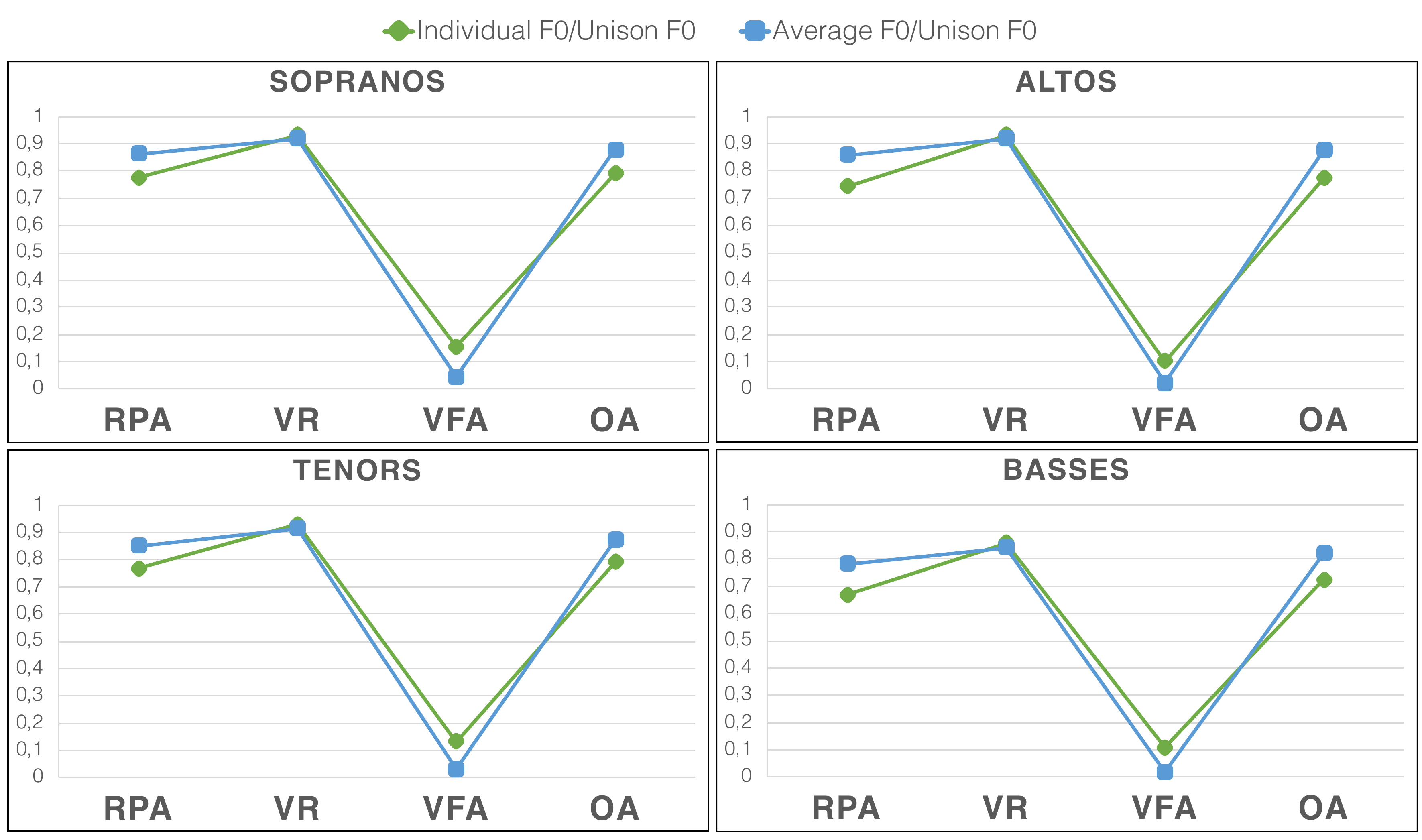}}}
 \caption{Resemblance of the estimated unison $EstF0_U$ estimation to each individual $GTF0_i$ contour (green) and the average (blue) using pitch evaluation metrics averaged across each choir section.}
 \label{fig:f0comparison}
\end{figure}

A summary of the results of the comparison between the fundamental frequency extracted by CREPE, $EstF0_U$, and the manually corrected fundamental frequency, $GTF0_i$ is illustrated in Figure~\ref{fig:f0comparison}, along with a comparison with the mean, $F0_m$. We observe that all sections follow the same pattern with similar metric values, and the unison F$0$ estimated by CREPE, $EstF0_U$, is closer to the average $F0_m$ ,than to the individual contours. In addition, all metrics improve when we compare the average F$0$ curve to the extracted F$0$ contour from the unison: RPA, VR and OA are higher in the blue plots, while VFA is lower. We can thus use the pitch estimated by CREPE, $EstF0_U$, as a representative of the mean single pitch contour perceived in a unison mixture ~\cite{Ternstrom91_PerceptualEvalVoice}.
\begin{figure}
 \centerline{
 \includegraphics[width=\columnwidth]{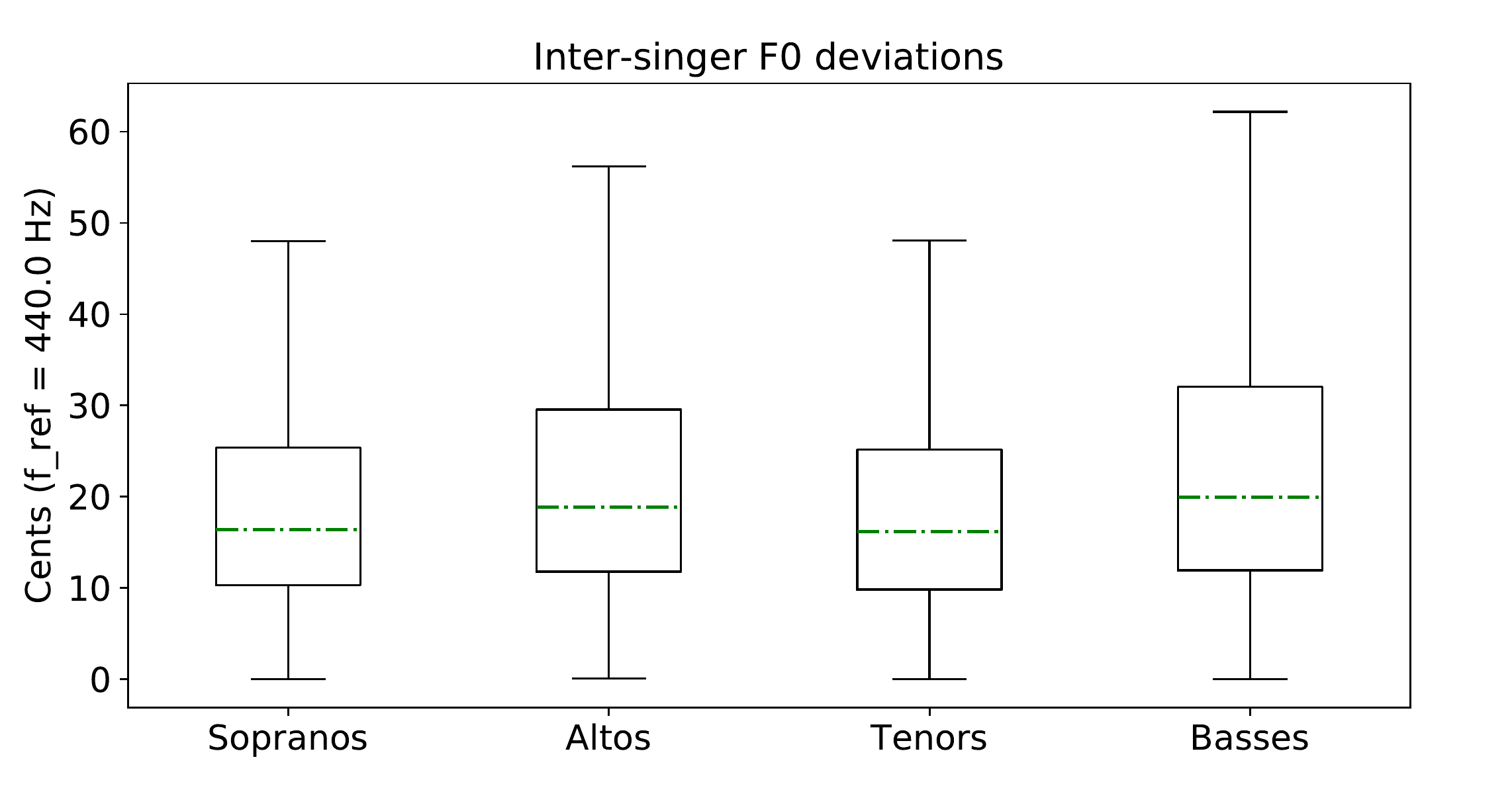}}
 \caption{Inter-singer deviations in cents averaged across the whole dataset for each choir section. Deviations are calculated using Equation~\ref{eq:deviation}.}
 \label{fig:deviation}
\end{figure}

The calculated $\Delta F0_s$ is shown in Figure~\ref{fig:deviation}. We observe an inter-singer deviation in the range of $\SIrange{0}{50}{cents}$, with a mean of around \SI{20}{cents}. This value, representing the inter-singer deviation in the unison mixtures, is comparable to the pitch dispersion studied by Cuesta et al.~\cite{CuestaGML18_ChoirIntonation_ICMPC}. While the methodology for modelling is different, these results are in accordance with their reported per-section pitch dispersion: larger in the bass section, smaller in the sopranos, and very similar for altos and tenors.

Table~\ref{tab:timing_deviations} shows the results of the timing analysis. We observe an average timing deviation of \SI{0.1}{\second} between the voices in the unison for all parts of the choir.
\begin{table}
 \begin{center}
 \begin{tabular}{cc}
 \toprule
  \bfseries Section & \bfseries Average Timing Deviation $\pm$\\
  & Standard Deviation\\
  \midrule
  Soprano & $0.134 \pm 0.039$ sec \\ \midrule
  Alto & $0.093 \pm 0.0024$ sec \\ \midrule
  Tenor  &  $0.100 \pm 0.021$ sec \\ \midrule
  Bass  & $0.124 \pm 0.021$ sec\\ 
  \bottomrule
 \end{tabular}
\end{center}
\caption{Timing deviations averaged across the CSD. These values measure the time span in which all singers in the unison transition from voiced to unvoiced, and vice-versa, averaged across all transitions in each song.}
 \label{tab:timing_deviations}
\end{table}
%
%
%
\section{Synthesis Methodology} \label{sec:method}
We present two synthesis models, Solo To Unison (STU) to synthesize voice clones for creating a unison mixture from a single voice input, and Unison to Solo (UTS) for synthesizing single voice prototype representing the melodic and linguistic content of a unison mixture input\footnote{Audio examples are provided as complementary material at \url{https://pc2752.github.io/unison_analysis_synthesis_examples/}and the source code with the trained models are available at \url{https://github.com/MTG/content_choral_separation}}. Similar to the work presented by Schnell et al~\cite{schnell2000synthesizing}, we decompose the input signal into the F$0$, harmonic spectral envelope, and aperiodicity envelope. However, instead of using Pitch Synchronous Overlap Add (PSOLA) methods, we utilize the WORLD vocoder~\cite{morise2016world}, which has been shown to be an effective vocoding system for singing voice synthesis~\cite{chandna2019wgansing, blaauw2017neural, blaauw2017neural}. Similar to~\cite{blaauw2017neural}, we use truncated frequency warping in the cepstral domain~\cite{TokudaKMI94_MelgeneralizedCA} to reduce the dimensions of the harmonic components from $1024$ to $60$ log Mel-Frequency Spectral Coefficients (MFSCs) with an all-pole filter with warping coefficient $\alpha=0.45$. In addition, we use bandwise aperiodic analysis to reduce the dimensionality of aperiodic features to $4$. For the rest of this paper, we refer to these $64$ features together as the spectral envelope.
\subsection{Unison to Solo (UTS)}
\begin{figure*}
 \centerline{
 \resizebox{0.85\textwidth}{!}{%
 \includegraphics{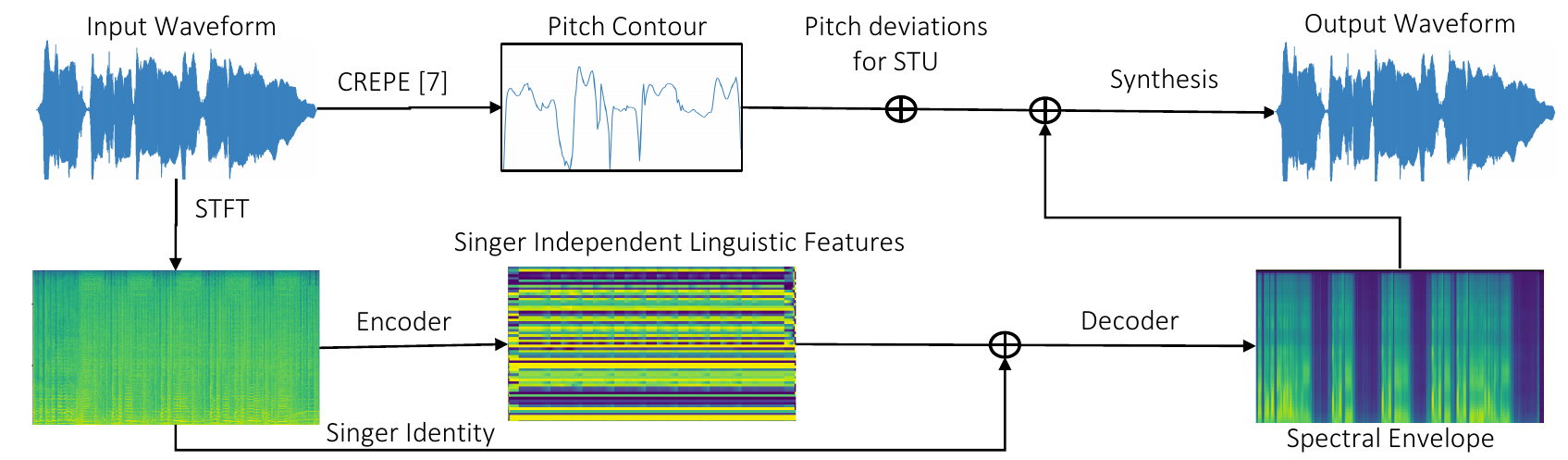}}}
 \caption{The synthesis framework. The magnitude part of the spectrogram of the input is passed through the encoder from \cite{chandna2020content} to extract linguistic content. For the UTS case, this is passed to the decoder along with a learned embedding to the SIN decoder~\cite{chandna2020content} to generate the spectral envelope. For STU, the linguistic features are passed to the SDN decoder along with a one-hot vector of the same gender as the input. 
 F$0$ is extracted from the input waveform via CREPE \cite{KimSLB18_CREPE_ICASSP}. Both the envelope and the F$0$ are used to synthesize the output voice. For the STU case, timing deviations are further added before summing with the input and normalizing.}
 \label{fig:UTS}
\end{figure*}

As shown in Figure~\ref{fig:UTS}, we first perform a short-time Fourier transform (STFT) to extract a spectrogram from the input waveform. The magnitude part of the spectrogram is passed through the encoder proposed in~\cite{chandna2020content} to extract a continuous representation of the linguistic features present in the unison mixture input. The linguistic features are decoded via the SIN network~\cite{chandna2020content} to generate the spectral envelope for vocal synthesis. This envelope is combined with the pitch contour output from CREPE~\cite{KimSLB18_CREPE_ICASSP}, $F0_U$ to synthesise the single voice prototype representing the unison mixture input. 
\subsection{Solo to Unison (STU)}
The analysis part of the STU case follows a similar methodology, as we extract the linguistic features and the $EstF0_i$ contour from the input \emph{a cappella} voice signal. To create voice clones with pitch and timing deviations, we add randomly sampled noise from a normal distribution with a mean of $0$ and a variable standard deviation, termed as $std$. This represents the inter-singer deviation, $\Delta F0_s$, and allows us to model the $F0_{i+1}$ of the clone as per Equation \ref{eq:model2}. Timing deviations are added by shifting the voiced portions of the input signal or the portions between tow blocks of silence of more than \SI{80}{\milli\second} by a variable amount, randomly sampled from a normal distribution of mean $0$ and standard deviation $ts$. The values of $std$ and $ts$ are based on our analysis of the Choral Singing Dataset presented in Section~\ref{sec:analysis:results}

Finally, for variations in timbre, we generate the spectral envelope of a variable number singers, $ns$, of the same gender as the input using the SDN network proposed in~\cite{chandna2020content}. This is based on our analysis presented in Section~\ref{sec:analysis:results}. There was no overlap between singers in the set used for training the synthesis model and the singers in the Choral Singing Dataset used for evaluation.
The various voice clones are added together and normalized in amplitude to produce the final unison output. We evaluate various combinations of $std$, $ts$, and $ns$ on their impact of the perception of unison. 

\subsection{Perceptual Evaluation Methodology}
\label{sec:evaluation}

We used subjective listening tests with low and high anchors, as modified versions of the MUSHRA-methodology~\cite{volker2018modifications} to evaluate subjective criteria of the synthesis produced by our analysis synthesis framework. 

While there are several aspects that could be evaluated, we focused on three keys aspects: adherence to melody, perception of unison, and audio quality. For each aspect, the participants were presented with \num{4} questions, one for each part of the SATB choir, and were asked to rate the test cases in the question on a continuous scale of $\SIrange{1}{5}{}$ with respect to a presented reference. The test case and references provided pertained to the the same section of the song and were between $\SIrange{7}{10}{\second}$ each. The parameters used for these tests are described below for each aspect.

\subsubsection{Adherence to melody and lyrics}
For this aspect, we wanted to see the similarity of the perceived pitch contour of the output for both the UTS and STU cases to that of a ground truth unison mixture. To this end, the reference provided to the participant was a ground truth unison sample made by summing the corresponding four individual singers of a part to form a unison mixture. This reference is referred to as \emph{REFU}. The participants were asked to rate test samples which included the single voice prototype of the unison as output by the UTS system, referred to as \emph{UTS}. In addition, we evaluated the output of STU with a pitch deviation with parameter $std$ set to \SI{50}{cents}, the acceptable limit of pitch deviations, as shown by our analysis in Section \ref{sec:analysis:results} and suggested by \cite{Ternstrom91_PerceptualEvalVoice}. Four singers were used for generating this test case, with parameter $ns$ set to \num{4}, and it is referred to as \emph{STU\_PS}. We also evaluated the output of the UTS system with both pitch and timing deviations with parameter $ts$ set to \SI{40}{\milli\second}. While our analysis in Section \ref{sec:analysis:results} suggests that higher values of $ts$ could have been used, we found that increasing the value beyond  \SI{40}{\milli\second} leads to a unacceptable level of degradation in output quality. We refer to this test case as \emph{STU\_PTS}. We also provided a lower anchor of a sample of the same length from another vocal part.

\subsubsection{Perception of unison}
Unison is a loosely defined perceptual aspect, the cognition of which we aim to study here. For this, we provide a reference of a ground truth unison sample created in the same manner as described above, \emph{REFU}. Given this, participants were asked to rate outputs from the STU system based on their similarity to the reference in terms of the perception of unison. In addition to the \emph{STU\_PTS} and \emph{STU\_PS} cases with pitch, timing and timbre variance, we also tested the case for just timing and singer variation, referred to as \emph{STU\_TS} and a case with just pitch and timing deviations, referred to as \emph{STU\_PT}, timbral changes were not done for the voice clones used for creating this test case. The \textit{a cappella} sample of a single singer singing the same example as the reference was provided as a lower anchor. 

\subsubsection{Audio Quality}
Audio quality is another subjective measure that is well defined in literature but not easily understood by non-expert participants. For the evaluation, we set an upper limit of audio quality to the resynthesis of a single voice recording with the WORLD vocoder \emph{REFS} and a lower limit to the resynthesis of a unison mixture with the same \emph{RESSYNTHU}. The test cases provided to the participants were the same as those provided for the adherence to melody case, except that the lower anchor was changed.

\subsection{Perceptual Evaluation Results}
\label{sec:results}
There were \num{17} participants in our evaluation, of which \num{10} had prior musical training. To account for inter-participant variance in subjective evaluation, the opinion score for each question was normalized over ratings for the reference and the lower anchor before calculating the mean opinion scores (MOS) and the standard deviations in opinion scores, presented in Table \ref{table:results}. 

\begin{table}
\centering
\begin{tabular}{l c c c c }

Test  &  Adherence To  & Unison  & Audio   \\ 
 Case & Melody & Perception & Quality \\ \toprule

UTS    & $3.6 \pm 0.93$& & $2.1 \pm 0.65$ \\

STU\_PS   & $3.3 \pm 0.83$ & $2.6 \pm 0.85$ & $2.8 \pm 0.45$ \\

STU\_PTS   & $2.9 \pm 1.14$& $3.2 \pm 0.96$ & $3.1 \pm 0.63$ \\

STU\_TS  & & $2.3 \pm 1.11$&  \\

STU\_PT  & & $3.0 \pm 1.23$&  \\

\end{tabular}
\caption{Mean Opinion Score (MOS) $\pm$ Standard Deviation for the perceptual listening tests across the test cases provided. The models shown corresdond to the Unison to Solo (UTS) model, the Solo to Unison with pitch, timing and singer variations, indicated by the addition of the letters P,T and S as suffixes to the abbreviation, respectively. The scores for each question were normalized by the responses to the upper and lower limits for the responses defined in section \ref{sec:evaluation}.}
\label{table:results}
\end{table}
The subjective nature of the perceptual aspects evaluated must be taken into account for the evaluation and the mean opinion scores are indicative of preferences rather than absolute measures of quantity. 
It can be observed that the perceived adherence to melody for the prototypical \emph{a cappella} voice synthesized by the UTS model has higher preference than the STU models, although a high variance is observed in the ratings for the same. The unison perception evaluation shows that the variations in either timing or pitch alone are not as preferred as variations in both aspects together. Timbre variations do not have as significant an effect on perception of unison as variances in timing and pitch. 
The evaluation of audio quality shows room for improvement in the synthesis of the voice signals. This can partly be attributed to the use of the WORLD vocoder~\cite{morise2016world} and we believe that this can be improved on in the future using recently proposed neural synthesis techniques.

\section{conclusions}
\label{sec:conclusions}
We have presented an analysis of the Choral Singing Dataset, building on the work presented in \cite{Ternstrom91_PerceptualEvalVoice}. 
In accordance with the analysis done by  \cite{CuestaGML18_ChoirIntonation_ICMPC}, we observe deviation in the range of $\SIrange{0}{50}{cents}$ between the F$0$ contours of the individual singers in the unison mixtures in the dataset. We further note an timing deviation of \SI{0.1}{\second} between singers in unison in the dataset. 

We then used this analysis along with recently proposed deep-learning based methodologies to present a synthesis system for a unison  mixture from a single voice input and a single  voice  prototype synthesis  representing  the melodic  and  linguistic  content  of  a  unison  mixture  input. Based on these systems, we were able to conduct a perceptual evaluation of the unison, further supporting the claim of \cite{Ternstrom91_PerceptualEvalVoice} that the a mixture of different voices singing in unison is perceived to have a single pitch. In addition, we found that pitch and timing deviations together are important for the perception of the unison, and that variations in either aspect alone is insufficient for such. However, timbre variations were not found to be as relevant.

We present this work as the first step into the analysis of an under-explored research area, hoping to fuel further discussion on the topic. While interesting from an academic standpoint, the systems we present also have several commercial applications such as creating a unison choral effect to be used in music production as well as for transposition and transcription, in conjunction with the work presented in~\cite{CuestaMG20_MultipitchVocalEnsembles_ISMIR}. We also plan to incorporate the presented work with~\cite{Peterman20}, for complete source separation for choral recordings. 
\section{acknowledgements}
The TITANX used for this research was donated by the NVIDIA Corporation. This work is partially supported by the Towards Richer Online Music Public-domain Archives (TROMPA H2020 770376) project. Helena Cuesta is supported by the FI Predoctoral Grant from AGAUR (Generalitat de Catalunya).

\label{sec:conclusions}
\bibliography{ReferencesMIR}

\end{document}